\documentclass{article}
\usepackage{amsmath}
\usepackage{amssymb}
\usepackage{times}
\newtheorem{Definition}{Definition}
\newtheorem{Lemma}{Lemma}

\newtheorem{Proposition}{Proposition}

 \newcommand{\begeq}{\begin{equation}}
\newcommand{\bea}{\begin{eqnarray}}
\newcommand{\eea}{\end{eqnarray}} \newcommand{\nn}{\nonumber}

\newcommand{\ca}{$C^*$-algebra}

 \newcommand{\til}{\tilde}
\newcommand{\raw}{\rightarrow}

\newcommand{\hraw}{\hookrightarrow}

\newcommand{\ot}{\otimes}

\newcommand{\x}{\times} 
 
\newcommand{\cin}{C^{\infty}}

\newcommand{\inv}{^{-1}}

\newcommand{\al}{\alpha} \newcommand{\bt}{\beta}
\newcommand{\gm}{\gamma} 
 
\newcommand{\ep}{\epsilon} \newcommand{\varep}{\varepsilon}

\newcommand{\lm}{\lambda} 
 \newcommand{\sg}{\sigma}
 \newcommand{\ta}{\tau} 
 
 \newcommand{\ps}{\psi}

\newcommand{\CQ}{{\mathcal Q}}

\newcommand{\C}{{\mathbb C}} 
 
 \newcommand{\R}{{\mathbb R}}
 \newcommand{\Z}{{\mathbb Z}}
  \makeatletter
\newskip\tempskip \def\endproof{{\parfillskip24\p@ plus\@ne
fil\@@par}\tempskip\prevdepth
\ifdim\lastskip=\z@\tempskip\z@\else\vskip-\lastskip
\ifdim\tempskip>4\p@ \tempskip.5\tempskip \else \tempskip\z@\fi\fi
\nobreak\vskip-\baselineskip\vskip-\tempskip\noindent\hbox
to\hsize{\hfill
$\blacksquare$}\par\vskip\tempskip\vskip\abovedisplayskip\@doendpe}
\makeatother \makeatletter
\newskip\tempskip \def\endiproof{{\parfillskip24\p@ plus\@ne
fil\@@par}\tempskip\prevdepth
\ifdim\lastskip=\z@\tempskip\z@\else\vskip-\lastskip
\ifdim\tempskip>4\p@ \tempskip.5\tempskip \else \tempskip\z@\fi\fi
\nobreak\vskip-\baselineskip\vskip-\tempskip\noindent\hbox
to\hsize{\hfill
$\Box$}\par\vskip\tempskip\vskip\abovedisplayskip\@doendpe}
\makeatother \newcommand{\enp}{\endproof}

\newcommand{\tind}{\mathrm{t}\mbox{\rm{-}}\mathrm{ind}}
\newcommand{\aind}{\mathrm{a}\mbox{\rm{-}}\mathrm{ind}}
\newcommand{\qind}{\mathrm{q}\mbox{\rm{-}}\mathrm{ind}}
\newcommand{\Ch}{\mathbb{C}[[\hbar]]}
\newcommand{\hh}{[[\hbar]]}

\newcommand{\BCC}{Baum--Connes conjecture}
 
\begin{document}
\pagestyle{plain}
\title{Quantization and the tangent groupoid\thanks{To appear in Proc.\ 
4th Operator Algebras International  Conference:
OPERATOR ALGEBRAS and MATHEMATICAL PHYSICS, 
                              Universitatea Ovidius, Constanta, Romania, July 2--7, 2001, eds. J. Cuntz et al.}}
\author{\textsc{N.P. Landsman}\thanks{Supported by a Fellowship from 
the Royal Netherlands Academy
of Arts and Sciences (KNAW).} \\
Korteweg--de Vries Institute for Mathematics\\
University of Amsterdam\\
Plantage Muidergracht 24\\
NL-1018 TV AMSTERDAM\\ THE NETHERLANDS\\
email \texttt{npl@science.uva.nl}}
\date{\today}
\maketitle 
\begin{abstract}  
This is a survey of the relationship between $C^*$-algebraic
deformation quantization and the tangent groupoid in noncommutative
geometry, emphasizing the role of index theory. We first explain how
$C^*$-algebraic versions of deformation quantization are related to
the bivariant E-theory of Connes and Higson. With this background, we
review how Weyl--Moyal quantization may be described using the tangent groupoid. Subsequently, we
explain how the Baum--Connes analytic assembly map in E-theory may be
seen as an equivariant version of Weyl--Moyal quantization.  Finally,
we expose Connes's tangent groupoid proof of the Atiyah--Singer index theorem.  \end{abstract}
\section{Introduction}
Quantization theory is concerned with the passage from classical to
quantum mechanics (or field theory), and vice versa. 
Dirac's famous early insight that the Poisson
bracket in classical mechanics is formally analogous to the commutator
in quantum mechanics was initially
implemented, in a mathematical context, in geometric quantization. This approach is
generally felt to be somewhat \textit{pass\'{e}}, although certain
techniques from it continue to play an important role.  What has
replaced geometric quantization is the idea of deformation
quantization, which emerged in the 1970s independently through the
work of Berezin \cite{Ber1,Ber2} and of Flato and his collaborators
\cite{BFFLS}.

 Here quantum
mechanics is seen as a deformation of classical mechanics, which
should be recovered as $\hbar\rightarrow 0$. Hence it is particularly
important to study quantum theory for a range of values of Planck's
``constant'' $\hbar$, and to control the classical limit. This aspect
was missing in geometric quantization, as was the idea that one should
start from Poisson manifolds, or, even more generally,
from Poisson algebras, rather than from symplectic manifolds.
Here a Poisson algebra is a commutative algebra $\til{A}$ over $\mathbb{C}$
equipped with a Lie bracket $\{\, ,\,\}$, such that
 for each $f\in \til{A}$ the map $g\mapsto \{f,g\}$
is a derivation of $\til{A}$ as a commutative algebra.
Seen in this light, the quickest definition of a Poisson manifold $P$ is that
the space $\til{A}=\cin(P)$ of smooth functions over it is a Poisson algebra
with respect to  pointwise multiplication.

The best-known approach to deformation quantization
is purely algebraic, and is known as formal deformation quantization
or star-product quantization. Here one works with formal power series
in $\hbar$; in particular, it is generally impossible to ascribe a numerical
value to Planck's constant. This approach was launched in
1978 \cite{BFFLS}, and has led to impressive existence
 and classification results so far. For
example, Fedosov proved by an explicit geometric construction that any
symplectic manifold can be quantized \cite{Fed1}, and Kontsevich, using
entirely different methods, extended this result to arbitrary Poisson
manifolds \cite{Kon1}. These results belong to the early phase of
formal deformation quantization, which has been reviewed by
Sternheimer \cite{Ste}.

Recently, the theory has been put on a new footing by Kontsevich and
Soibelman, who use a high-powered description of general deformation
theory in terms of operads \cite{Kon2,KS}. Their approach uncovers
unexpected and fascinating links between deformation quantization, the theory
of motives,  and the so-called Grothendieck--Teichm\"{u}ller group
in algebraic geometry. This illustrates the phenomenon that 
despite its original motivation, formal
deformation quantization is taking a path that is increasingly remote from physics.

The link between operator algebras and quantum physics has been close 
ever since von Neumann's foundational work in both areas. It should,
therefore, be no surprise that $C^*$-algebras provide a language for describing 
deformation quantization that is interesting for both mathematics and physics.
The physical interest in the \ca ic approach lies partly in the fact that $\hbar$
is now a real number rather than a formal parameter, so that 
one can study the
limit $\hbar\rightarrow 0$ in a precise, analytic way, and partly
in the possibility of explicitly describing most known examples of quantization
as it is used in physics. Mathematically, it turns out that \ca ic deformation
quantization sheds light on many interesting examples in noncommutative geometry.
(In this paper, we always mean ``noncommutative geometry'' in the sense of Connes
\cite{Con}.
There are constructions involving homotopic algebra and ``$\infty$-structures''
that go under this name as well, and which are actually closely related to
\textit{formal} deformation quantization; see \cite{TT} for a
representative paper.) 

The $C^*$-algebraic approach to deformation quantization was initiated
in 1989 by  Rieffel \cite{Rie1}, who observed that a number of examples of
quantization could be described by continuous fields of $C^*$-algebras
in a natural and attractive way. As
indicated above, some of his examples involve quantization as physicists know and love it, like
Weyl--Moyal quantization and related constructions (see, in particular, \cite{Rie2} 
for a survey), while others relate to noncommutative geometry.
In the latter category, Rieffel's discovery that the familiar 
noncommutative tori can be seen
as deformation quantizations of ordinary symplectic tori stands out
\cite{Rie1,Rie3}. (Noncommutative tori actually do have 
potential physical relevance  through string theory \cite{CDS}.)

We refer to \cite{Lan,Rie2} for surveys of the  starting period of $C^*$-algebraic
deformation quantization, including references
up to 1998.  Later work that is relevant to noncommutative geometry
includes \cite{LGCA,LR}, which will be recalled below, as well as
\cite{NN}.  Very recently, Cadet \cite{Cad} showed that the Connes--Landi
noncommutative four-spheres \cite{CL} fall into this context. The
general picture of $C^*$-algebraic deformation quantization that
emerges from the literature so far is that it is rich in examples and
poor in existence and classification theorems; compare this with the
formal case!

We now outline the contents of the remainder of this paper; the key concept
unifying what follows is Connes's tangent groupoid \cite{Con,HS}.
It is clear from its very definition that the bivariant E-theory of Connes and Higson 
\cite{Bla,Con,CH} should be closely related to $C^*$-algebraic deformation quantization
as formulated by Rieffel \cite{Nag,Ros}. In Section \ref{s2} we sketch
a direct route from formal deformation quantization to asymptotic
morphisms and E-theory, which entices a generalization of Rieffel's
$C^*$-algebraic axioms.  In Section \ref{s3} we sketch
an approach to Weyl--Moyal quantization that is based on a powerful
lemma, which in Section \ref{s4} we show  to underlie  the Baum--Connes conjecture \cite{BC,BCH} in E-theory as formulated in \cite{Con}. 
  Since the Baum--Connes conjecture is an issue in index
  theory, our discussion is intended as a minor contribution to the growing literature on the intimate relationship between deformation quantization, K-theory, and index theory.  In the purely algebraic setting, powerful new
  results have been achieved in this direction
\cite{Fed1,Fed2,Fed3,NT1,NT2,NT3}, whereas  \ca ic quantization-oriented
methods so far have mainly led to new proofs of known results.  In
the latter spirit, Section \ref{s5} contains  an exposition of   Connes's tangent groupoid proof of the Atiyah--Singer index theorem \cite{Con}. 

Throughout this paper we use the following convention. 
$G$ is a Lie groupoid over $G^{(0)}$, with associated convolution
\ca s $C^*(G)$ and $C^*_r(G)$ \cite{Con}. We write $K^*(G)$ for $K_*(C^*(G))$, and similarly  $K^*_r(G)=K_*(C^*_r(G))$.
This is consistent with the usual identification $K^*(X)=K_*(C_0(X))$,
for when a locally compact groupoid $G$ is a space $X$ (in that
$G=G^{(0)}=X$ with trivial operations), one has $C^*(X)=C_0(X)$. 

\bigskip

\textbf{Acknowledgement} The author is indebted to Erik van Erp for many conversations
about index theory. He also wishes to thank the organizers of the Constanta meeting
for putting together such an interesting and pleasant conference.
\section{From deformation quantization to E-theory}\label{s2}
In formal deformation quantization one defines a star-product $*$ on a
unital Poisson algebra $\til{A}$ as an associative product on the ring
$\til{A}\hh$ of formal power series in one variable with coefficients
in $\til{A}$ \cite{BFFLS}. Such a product is evidently determined by
its value on $\til{A}$. Writing $f*g=\sum_n \hbar^{2k} C_n(f,g)$, where
$f,g\in \til{A}$, one requires that $C_0(f,g)=fg$ and
$C_1(f,g)-C_1(g,f)=i\{f,g\}$. Heuristically, one may think of the
restriction of the star-product $*$ to $\til{A}$ as a family of
associative products $*_{\hbar}$ on $\til{A}$.
 
Rieffel's original definition of \ca ic deformation quantization \cite{Rie1}
was motivated by this interpretation. He defined a ``strict''quantization of a given 
Poisson manifold $P$ as a family $(A_{\hbar})_{\hbar\in I}$
of \ca s, equipped with the structure of a continuous field, with the feature that
each fiber algebra $A_{\hbar}$ is the completion of a single (i.e., $\hbar$-independent)
Poisson algebra $\til{A}_0$ that is densely contained in
the commutative \ca\ $A_0=C_0(P)$, equipped with  a ``deformed'' (i.e., $\hbar$-dependent)
 product $*_{\hbar}$,  involution $\mbox{}^{*_{\hbar}}$, and  norm
$\|\cdot \|_{\hbar}$. Here one generically takes $\hbar\in I=[0,1]$, although more general 
base spaces of the continuous field are occasionally used 
(as long as the base contains 0 as an accumulation point).

Consequently, one has canonical ``quantization'' maps 
$Q_{\hbar}:\til{A}_0\raw A_{\hbar}$ given by
$Q_{\hbar}(f)=f$, seen as an element of $A_{\hbar}$, and
for each $f\in\til{A}_0$
the map $\hbar\mapsto Q_{\hbar}(f)$ 
defines a canonical
section of the field. By construction, one then has 
\begin{equation}
Q_{\hbar}(f)*_{\hbar}Q_{\hbar}(g)=Q_{\hbar}(f *_{\hbar}g) \label{R1}
\end{equation} for all $f,g\in \til{A}_0$. 
Hence Rieffel was able to formulate Dirac's insight mentioned earlier in an asymptotic way
by means of the axiom
\begin{equation}
\lim_{\hbar\rightarrow 0} 
\|\frac{i}{\hbar}[Q_{\hbar}(f),Q_{\hbar}(g)]-Q_{\hbar}(\{f,g\})\|_{\hbar} =0 \label{Dirac}
\end{equation}
for all $f,g\in \til{A}_0$. Here $[\, ,\,]$ is the commutator with respect to
$*_{\hbar}$.

In examples related to Berezin--Toeplitz quantization, however, 
continuous fields of \ca s and quantization maps $Q_{\hbar}$ occur which do not have 
the feature that $Q_{\hbar}(f)Q_{\hbar}(g)$ is the $Q_{\hbar}$ of something,
contra (\ref{R1}); see \cite{Lan} and references therein.
This called for a more general definition of \ca ic deformation quantization
\cite{NPL93,Lan,Sheu}, whose relationship
with formal deformation quantization was rather obscure. We now remove this deficiency.

The algebra $\til{A}\hh$ used in the formal setting is
 a $\Ch$ algebra, in the sense that there is an injective  ring homomorphism from
$\Ch$ into the center of $\til{A}\hh$; cf.\ \cite[p.\ 121]{Lang}. 
Now the \ca ic analogue of such an algebra is a so-called $C(I)$ \ca. Recall that,
for a compact Hausdorff space $X$, a $C(X)$ \ca\ is a \ca\ $A$
with a unital embedding of $C(X)$ in the center of its multiplier algebra \cite{Kas}.
The structure of $C(X)$ \ca s is  as follows \cite{Nil}.

A field of \ca s is a triple $(X, \{A_x\}_{x\in X}, A)$,
where $\{A_x\}_{x\in X}$ is some family of \ca s indexed by $X$, and
$A$ is a family of sections (that is, maps $f:X\raw \coprod_{x\in X}A_x$ for which
$f(x)\in A_x$) that is \textit{i)}
 a \ca\ under pointwise operations and the natural norm 
$\| f\|=\sup_{x\in X} \| f(x)\|_{A_x}$, \textit{ii)}
 closed under multiplication by $C(X)$, and \textit{iii)}
full, in that for each $x\in X$ one has $\{f(x)\mid f\in A\}=A_x$.
The field is said to be continuous when for each $f\in A$ the function $x\mapsto \| f(x)\|$
is in $C(X)$ (this is equivalent to the corresponding definition of Dixmier \cite{Dix}; cf.\ 
\cite{KW}).  The field is upper semicontinuous when for each $f\in A$
and each $\varep>0$ the set $\{x\in X\mid \| f(x)\|\geq\varep\}$ is compact.

Thm.\ 2.3 in \cite{Nil} now states that  a $C(X)$ \ca\  $A$ defines a unique
upper semicontinuous field of \ca s $(X, \{A_x=A/C(X,x)A\}_{x\in X}, A)$.
Here $C(X,x)=\{f\in C(X)\mid f(x)=0\}$, and,
with slight abuse of notation, $a \in A$ is identified with the section 
\begin{equation}
a:x\mapsto\pi_x(a),\label{sec}
\end{equation}
 where $\pi_x:A\raw A_x$ is the canonical projection.
Moreover,  a $C(X)$ \ca\  $A$ defines a continuous field of \ca s whenever
the map $x\mapsto \| \pi_x(a)\|$ is lower semicontinuous 
(and hence continuous) for each $a\in A$ \cite{Blan}.

We return to deformation quantization. In the formal setting, 
given a Poisson algebra $\til{A}$ one could look at general
$\Ch$ algebras $A$ with the property that $A/\hbar A\cong \til{A}$, rather than
narrowing the discussion to the free $\Ch$ modules $\til{A}\hh$. 
This motivates the following definition in the analytic context. 
As in Rieffel's discussion, we start from a Poisson manifold instead of
a Poisson algebra. 
\begin{Definition}\label{gsq}
A \ca ic quantization of a Poisson manifold $P$ is a 
$C(I)$ \ca\ $A$ such that
\begin{enumerate}
\item For each $a\in A$, the function $\hbar\mapsto \| \pi_{\hbar}(a)\|$ from $I$ to $\R^+$
is  lower semicontinuous (and hence continuous);   
\item
One has $A_0=A/C(I,0)A\cong C_0(P)$ as \ca s;
\item There is a 
Poisson algebra $\til{A}_0$ that is densely contained in
$C_0(P)$, and,  identifying $A_0$ and $C_0(P)$,
there is a cross-section $Q:\til{A}_0\raw A$ of $\pi_0$,  
such that (\ref{Dirac}) holds for 
$Q_{\hbar}=\pi_{\hbar}\circ Q$.
 \end{enumerate}
\end{Definition}

This definition (with evident modifications when $I=[0,1]$ is replaced by
a more general index set) seems to cover all known examples.
It follows from the discussion above that, due to the first condition, $A$
is automatically the section algebra of a continuous field. 
Let us now assume that this field is trivial
away from $\hbar=0$. This means  by definition that $A_{\hbar}=B$ for all $\hbar\in (0,1]$,
and that, under the identification (\ref{sec}), one has a short exact sequence 
\begin{equation}
0\raw CB\raw A\raw A_0\raw 0. \label{SES}
\end{equation}
Here the so-called cone  $CB=C_0((0,1],B)$ appears.
(Strictly speaking, the fields in  our examples are merely isomorphic
to those of this form, but there is always a canonical trivialization.)

In this situation, one obtains a homomorphism $\CQ_*$ from $K_*(A_0)$
to $K_*(B)$, as follows.  Since the cone $CB$ is contractible, and
therefore has trivial K-theory, the periodic six-term sequence shows
that
\begin{equation}
\pi_0: K_*(A)\raw K_*(A_0) \label{Kiso}
\end{equation}
is an isomorphism. (In fact, Bott periodicity is not needed to infer that $\pi_0$ is
invertible; the long exact sequence of K-theory with an ad-hoc argument will do.)
Here, with abuse of notation, $\pi_0$ stands for the image of 
the $\mbox{}^*$-homomorphism $\pi_0:A\raw A_0$ under the K-functor.
(See  \cite{Ros} for the analogous result $K_0(\til{A}\hh)\cong K_0(\til{A})$ 
in formal deformation quantization.)   
The K-theory map defined by the continuous field is then simply
\begin{equation}
\CQ_*=\pi_1\circ\pi_0\inv: K_*(A_0)\raw K_*(B). \label{Kmap}
\end{equation}

This map may be described more explicitly, whether or not $A_0$
is commutative, as follows \cite{ENN1}. 
Denote the unitization of a \ca\ $C$ without unit by $C^+$, and assume
for simplicity that neither $A_0$ nor $B$ (and hence $A$)
is unital (this is indeed the case in all our examples).
Firstly, for any $n\in\mathbb{N}$, the \ca\ $M_n(A^+)$ 
of $n\x n$ matrices over $A^+$ is again a $C(I)$ \ca, and 
a nontrivial argument shows that it even defines a continuous field whenever $A$ does
\cite{ENN1}. The fiber algebras of this field are evidently $M_n(A_0^+)$
at $\hbar=0$ and  $M_n(B^+)$ at $\hbar\in(0,1]$. Now let $[p]-[q]\in K_0(A_0)$, where
$p,q$ are projections in $M_n(A_0^+)$. Extend $p$ and $q$ to continuous
sections $\hbar\mapsto p_{\hbar}$ etc.\ of the field $M_n(A^+)$, and finally put
\begin{equation}
\CQ_0([p]-[q])=[p_1]-[q_1], \label{defCQ}
\end{equation}
which lies in $K_0(B)$ as desired. This is independent of all choices.
Of course, the suffix 1 may be replaced by $\hbar$ for any
$\hbar\in(0,1]$. To construct $\CQ_1$, one works with suspensions as
appropriate.

The passage to E-theory is well known \cite{Bla,Con,CH,Nag,Ros}, as
follows.  Any cross-section $Q:A_0\raw A$ of $\pi_0$ defines an
asymptotic morphism $(Q_{\hbar})_{\hbar\in I}$ from $A_0$ to $B$ by
$Q_{\hbar}=\pi_{\hbar}\circ Q:A_0\raw B$, and all such $Q$ define
homotopic asymptotic morphisms.  Thus a deformation quantization
defines an element of $E(A_0,B)$, and therefore a homomorphism from
$K_*(A_0)$ to $K_*(B)$.  This  homomorphism is precisely $\CQ_*$, which in the context of asymptotic morphisms has an explicit description, too
\cite{Hig}: extend the $Q_{\hbar}$ to maps $Q_{\hbar}^n:M_n(A_0^+)\raw
M_n(B^+)$ in the obvious way, and find continuous families of
projections $(p_{\hbar})_{\hbar\in(0,1]}$ in $M_n(B^+)$ etc.\ such
that
\begin{equation}
 \lim_{\hbar\raw 0}  \| Q_{\hbar}^n(p)-p_{\hbar}\| =0.
\end{equation}
  Then use (\ref{defCQ}) as above.

In fact, it is sufficient if $Q$ is defined on a dense subspace
 $\til{A}_0$ of $A_0$, as in Definition
\ref{gsq}. The corresponding $\mbox{}^*$-homomorphism
from $\til{A}_0$ to $C_b((0,1],B)/CB$ can be extended to
$A_0$ by continuity, and this extension may subsequently be lifted to
an asymptotic morphism from $A_0$ to $B$, which on $\til{A}_0$ 
is equivalent to the original one.

 By the same argument, one may start from a definition
of quantization directly in terms of maps $Q_{\hbar}: \til{A}_0\raw B$,
as in \cite{NPL93,Lan}, and arrive at E-theory classes, but in the examples
below it will be the $C(I)$ \ca s rather than their associated continuous fields
or quantization maps that are canonically given. 
A \ca ic quantization has more structure than an asymptotic morphism
in E-theory,
 in that in the latter the maps $Q$ are completely
arbitrary, whereas in the former they relate to the Poisson structure on
$A_0$, and have to be chosen with care. This is clear from condition 3 in
Definition \ref{gsq}, on which the transition from deformation quantization
to E-theory does not depend.
\section{Weyl--Moyal quantization}\label{s3} 
The first example to consider in any version of quantization theory is
the Weyl--Moyal quantization of $T^*(\mathbb{R}^n)$, or more
generally, of $T^*(M)$, where $M$ is a Riemannian manifold. In the
formal setting this is handled for $\R^n$ in \cite{BFFLS} and for
general $M$ in \cite{DWL,Pflaum}; for the \ca ic formalism we refer to
\cite{Rie2} and \cite{NPL93,Lan}, respectively. In the context of
noncommutative geometry and the Baum--Connes conjecture, the ``royal
path'' towards Weyl--Moyal quantization \cite{CCFGRV,Lan,LGCA} is
formulated in terms of Connes's tangent groupoid (cf.\ \S II.5 in
\cite{Con}), as follows.\footnote{After circulation of this paper as a
preprint I heard from Alejandro Rivero that Connes himself suggested
this formulation at Les Houches 1995.}  An immersion $M\hraw N$ of
manifolds defines a manifold with boundary
\begin{equation}
G_{M\hraw N}= \{0\}\x\nu(M)\cup (0,1]\x N, \label{GMN}
\end{equation}
where $\nu(M)$ is the normal bundle of the embedding. 
The smooth structure on this space  was first defined in \cite{HS}. 
If $N=M\x M$ and the embedding is the diagonal map $x\mapsto (x,x)$,
the ensuing manifold $G_{M\hraw M\x M}$, denoted simply by $G_M$ in what
follows, is a Lie groupoid over $G_M^{(0)}=I\x M$ in the following way.
The fiber at $\hbar=0$ is $\nu(M)=T(M)$, which is a groupoid over $M$ under
the canonical bundle projection and addition in each $T_x(M)$.
The fiber at any $\hbar\in(0,1]$ is the pair groupoid $M\x M$ over $M$.    
The total space $G_M$, then, is a groupoid with respect to fiberwise operations. This Lie groupoid is the tangent groupoid of $M$.
See also  \cite{Lan,Pat}. 

It is quite obvious that $A=C^*(G_M)$ is a $C(I)$ \ca, with associated fiber algebras
\begin{eqnarray}
A_0 & = & C_0(T^*(M)) ;\nn \\
A_{\hbar} & = & B_0(L^2(M))\:\: \forall\hbar\in(0,1], \label{CF1}
\end{eqnarray}
where $B_0(H)$ is the \ca\ of compact operators on $H$.
The continuity of this field may be established in many ways
(see \cite{Lan,Rie2} and references therein), but in the context of this
paper the most appropriate approach is to use the following lemma,
due to Blanchard and Skandalis (but apparently first published in 
\cite{LR}, which is partly based on Ramazan's thesis \cite{Ram}). This lemma
generalizes a corresponding result of Rieffel \cite{Rie0} from groups to groupoids. 
We only state and need the smooth case.
\begin{Lemma} \label{Ramlem}
Let $H$  be a Lie groupoid fibered over a 
manifold $X$ by a smooth surjective submersion $\pi :H\raw
X$ (both $H$ and $X$ may be manifolds with boundary).  
Suppose that $\pi(x)=\pi(s(x))=\pi(r(x))$ (where $s$ and $r$
are the source and the range projections in $H$); in that case, each $H_x=\pi\inv(x)$ is a Lie
subroupoid of $H$, and $H$ is a bundle of Lie groupoids over $X$ with fibers $H_x$ and
pointwise operations. 

Then $(X, \{C^*(H_x)\}_{x\in X}, C^*(H))$ is a field of \ca s, which is continuous at all points $x$ where $H_x$ is amenable. The same statement holds if $C^*(H_x)$ and $C^*(H)$ are replaced by $C_r^*(H_x)$ and $C_r^*(H)$, respectively.
\end{Lemma}

See \cite{AR} for the theory of amenable groupoids.
Applied to the tangent groupoid $H=G_M$, where $X=I$,  this lemma 
 proves continuity of the
field (\ref{CF1}), since the groupoid $H_0=T(M)$ is commutative and therefore amenable, and $H_{\hbar\neq 0}=M\x M$ is amenable as well.
In fact, equipping the  cotangent bundle $T^*(M)$ with the canonical
Poisson structure, all of Definition \ref{gsq} holds \cite{Lan,Rie2};  
the quantization maps $Q_{\hbar}$ 
may be given by Weyl--Moyal quantization with respect to a Riemannian structure on $M$. 
\section{The Baum--Connes conjecture in E-theory}\label{s4}
The Baum--Connes conjecture \cite{BC,BCH,Con} is an important issue 
in noncommutative geometry; see \cite{Val} for a recent
overview focusing on discrete groups, and cf.\ 
\cite{Tu} for a survey of the
situation for groupoids. The purpose of this section is to show how 
 Connes's  E-theoretic description of the analytic assembly map \cite[Ch.\ II]{Con} approach fits into the formalism of the previous sections, simultaneously inserting some details omitted in section II.10.$\alpha$ of
\cite{Con}. We will use the notation of \cite{Con}.

 Recall \cite{Con,Mac,Lan} that a (right) $G$ space $P$ is a
smooth map $P\stackrel{\al}{\raw}G^{(0)}$ along with a map $P\x_{\al}
G\raw P$, $(p,\gm)\mapsto p\gm$ (where $\al(p)=r(\gm)$), such that
$(p\gm_1)\gm_2=p(\gm_1\gm_2)$ whenever defined, $p\al(p)=p$ for all
$p$, and $\al(p\gm)=s(\gm)$. The action is called proper when $\al$ is
a surjective submersion and the map $P\x_{\al} G\raw P\x P$,
$(p,\gm)\mapsto (p,p\gm)$ is proper (in that the inverse images of
compact sets are compact).

The following construction is crucial for what follows. Let a $G$
space $H$ be a Lie groupoid itself, and suppose the base map
$H\stackrel{\al}{\raw}G^{(0)}$ is a surjective submersion that
satisfies $\al\circ s_H=\al\circ r_H=\al$ as well as the condition that,  
for each $\gm\in G$, the map $\al\inv(r(\gm))\raw \al\inv(s(\gm))$,
$h\mapsto h\gm$, is an isomorphism of Lie groupoids (note that for each
$u\in G^{(0)}$, $\al\inv(u)$ is a Lie groupoid over $\al\inv(u)\cap H^{(0)}$).
In particular, one has $(h_1h_2)\gm=(h_1\gm)(h_2\gm)$ whenever defined.
  
Under these conditions, one may define a Lie groupoid $H\rtimes G$,
 called the semidirect product
of $H$ and $G$ (see
 \cite{AR} for the locally compact case and \cite{Mac} (2nd ed.)
for the smooth case). The total space of $H\rtimes G$ is $H\x_{\al} G$,
the base space of units $(H\rtimes G)^{(0)}$is $H^{(0)}$, the source and range
maps are 
\begin{eqnarray}
s(h,\gm) & = & s_H(h)\gm; \nn \\
r(h,\gm) & = & r_H(h), \label{sr}
\end{eqnarray}
respectively, 
the inverse is $(h,\gm)\inv =(h\inv\gm,\gm\inv)$ (note that
one automatically has $\al(h\inv)=\al(h)$, so that this element is well defined), 
and multiplication is given by $(h_1,\gm_1)(h_2\gm_1,\gm_2)=(h_1h_2,\gm_1\gm_2)$,
defined whenever the product on the right-hand side exists (this follows from
the automatic $G$ equivariance of $s_H$ and $r_H$). Well-known special 
cases of this construction occur when $H$ is a space and $G$ is a groupoid,
so that $H\rtimes G$ is a groupoid over $H$, and when $G$ and $H$ are both groups,
so that $H\rtimes G$ is the usual semidirect product of groups.

In the context of the \BCC, the key application of this construction
is as follows \cite{Con}.  Let $P$ be a proper $G$ space. One may define three
Lie groupoids, all over $P$. 
\begin{enumerate}
\item
The tangent bundle $T_G(P)$ of $P$
along $\al$ (i.e., $\ker(\al_*)$, where $\al_*:T(P)\raw T(G^{(0)})$ is the
derivative of $\al$) is a $G$ space, with base map $\al_0(\xi_p)
=\al(p)$ (where $\xi_p\in T_G(P)_p$) and with
the obvious push-forward action. If $T_G(P)$ is seen as a Lie groupoid
over $P$ by inheriting the Lie groupoid structure from $T(P)$
(see Section \ref{s3}), one may define the semidirect product groupoid
$T_G(P)\rtimes G$ over $P$.
\item The fibered product $P\x_{\al} P$ is a $G$ space under the 
base map $\al_1(p,q)=\al(p)=\al(q)$ and the diagonal
action $(p,q)\gm=(p\gm,q\gm)$. Moreover, $P\x_{\al} P$ inherits
a Lie groupoid structure from the pair groupoid $P\x P$ over $P$, 
becoming a Lie groupoid over $P$. Hence one has the semidirect
product groupoid $(P\x_{\al} P)\rtimes G$ over $P$.
\item The tangent groupoid $G_P$ associated to $P$ has a Lie subgroupoid
$G_P'$ over $I\x P$ that by definition contains all points $(\hbar=0,\xi_p)$
of $G_P$ whose $\xi_p$ lies in $T_G(P)$, and all points $(\hbar>0,p,q)$
for which $\al(p)=\al(q)$. It is clear that $G_P'$ is a bundle of groupoids
over $I$, whose  fiber at $\hbar=0$ is $T_G(P)$, and whose fiber at any 
$\hbar\in(0,1]$ is $P\x_{\al} P$. Combining the $G$ actions defined in 
the preceding two items, there is an obvious fiberwise $G$ action on $G_P'$
with respect to a base map $\til{\al}(\hbar,\cdot)=\al_{\hbar}(\cdot)$,
where $\al_{\hbar}=\al_1$ for $\hbar\in(0,1]$. This action
is smooth, so that one obtains a semidirect Lie groupoid $G_P'\rtimes G$ 
over $I\x P$.
\end{enumerate} 

The following two propositions provide the technical underpinning for \S II.10.$\al$
in \cite{Con}.
\begin{Proposition}\label{conpin1}
If $P$ is a proper $G$ space, then $C^*(G_P'\rtimes G)$ is the 
\ca\ of sections $A$ of a continuous field of \ca s over $I$ with fibers
\begin{eqnarray}
A_0 & = & C^*(T_G(P)\rtimes G) ; \nn\\
A_{\hbar} & = & C^*((P\x_{\al} P)\rtimes G)\:\: \forall\hbar\in(0,1]. \label{CF3}
\end{eqnarray}
This field is trivial away from $\hbar=0$. The same is true if all
groupoid \ca s  are replaced by their reduced counterparts.
\end{Proposition}

\textit{Proof.} It is obvious that $G_P'\rtimes G$ is a bundle
of groupoids over $I$, whose  fiber at $\hbar=0$ is $T_G(P)\rtimes G$, and whose 
fiber at any  $\hbar\in(0,1]$ is $(P\x_{\al} P)\rtimes G$. Since the corresponding field of \ca s is obviously trivial away from $\hbar=0$, 
it is continuous at all $\hbar\in(0,1]$. 
If we can show that $T_G(P)\rtimes G$ is an amenable groupoid,
 Lemma \ref{Ramlem} proves continuity at $\hbar=0$ as well. 

To do so, we use Cor.\ 5.2.31 in \cite{AR}, which states that a (Lie) groupoid
$H$ is amenable iff the associated principal groupoid (that is, the image
of the map $H\raw H^{(0)}\x  H^{(0)}$, $h\mapsto (r(h),s(h))$) is amenable
and all stability groups of $H$ are amenable. As to the first condition, the
principal groupoid of $T_G(P)\rtimes G$ is the equivalence relation
on $P$ defined by $p\sim q$ when $q=p\gm$ for some $\gm\in G$. 
This is indeed amenable, because this equivalence relation is at the same
time the principal groupoid of $P\rtimes G$ (over $P$), 
which is proper (hence amenable) because $P$ is a proper $G$ space.
As to the second condition, the stability group of $p\in P$ in $T_G(P)\rtimes G$
is $T_G(P)_p\rtimes G_p$, where $G_p$ is the stability group of $p\in P$
in $P\rtimes G$. The latter is compact by the properness of the $G$ action,
so that $T_G(P)_p\rtimes G_p$ is amenable as the semidirect product of two amenable
groups. \enp

When $G$ is trivial, the continuous field of this proposition is, of course,
the one defined by the tangent groupoid of $P$, which coincides with the field
defined by the Weyl--Moyal quantization of the cotangent bundle $T^*(P)$; 
see Section \ref{s3}. The general case is a $G$ equivariant version of quantization,
which cannot really be interpreted in terms of quantization, because the fiber algebra
at $\hbar=0$ is no longer commutative.
\begin{Proposition}\label{conpin2}
The \ca s $C^*((P\x_{\al} P)\rtimes G)$ and $C^*(G)$ are (strongly) Morita
equivalent, as are the corresponding reduced \ca s.
\end{Proposition}

\textit{Proof.} It is easily checked that the map
$(p,q,\gm)\mapsto\gm$ from $(P\x_{\al} P)\rtimes G$ to $G$ is an equivalence
of categories. Since this map is smooth, it follows from Cor.\ 4.23 in
\cite{OBWF} that  $(P\x_{\al} P)\rtimes G$ and $G$ are Morita equivalent
as Lie groupoids (and hence as locally compact groupoids with Haar system).
The proposition then follows from Thm.\ 2.8 in \cite{MRW}.
\enp

By (\ref{Kmap}), the continuous field of Proposition \ref{conpin1}
yields a map 
\begin{equation} \CQ_*: K^*(T_G(P)\rtimes G)\raw
K^*((P\x_{\al} P)\rtimes G).
\end{equation}
 By Proposition \ref{conpin2} and
the fact that the K-theories of Morita equivalent \ca s are
isomorphic, this map equally well takes values in $K^*(G)$, and
hence, by the K-theory push-forward of the
canonical projection $C^*(G)\raw C_r^*(G)$, in $K^*_r(G)$.

Now suppose that the classifying space $\underline{E}G$ for proper $G$
actions is a smooth manifold (which is true, for example, when $G$ is
a connected Lie group \cite[\S II.10.$\bt$]{Con}, 
or when $G$ is the tangent groupoid of a manifold). This means that,
 up to homotopy, there is a unique smooth
$G$-equivariant map from any proper $G$ manifold to $\underline{E}G$.  
In that case, one may put $P=\underline{E}G$ in the above formalism, and,
writing 
\begin{equation}
K^*_{\mathrm{top}}(G)= K^*(T_G(\underline{E}G)\rtimes G), \label{Ktop}
\end{equation}
one obtains a map
\begin{equation}
\mu:K^*_{\mathrm{top}}(G)\raw K^*_r(G).
\end{equation}
This is the analytic assembly map in E-theory as defined by Connes. 
In  general, the definition of $K^*_{\mathrm{top}}(G)$ is
more involved, but the analytic assembly map is constructed using
the  above construction in a crucial way.  
The Baum--Connes conjecture (without coefficients) in E-theory
states that $\mu$ be an isomorphism. It remains to be seen how this
relates to the Baum--Connes conjecture for groupoids in KK-theory
\cite{Tu}, which is a priori stronger even if the assembly maps 
turn out to be the same. For further comments cf.\ the end of the next section. 
\section{The Atiyah--Singer index theorem}\label{s5}
We now use the ideas in the preceding sections to sketch two proofs of
the Atiyah--Singer index theorem. We refer to
\cite{AS1,AS3,LM} for the necessary background. Throughout this section, $M$ is a compact
manifold.  Atiyah and Singer \cite{AS1} define two maps, $\tind$ and
$\aind$, from $K^0(T^*(M))$ to $\Z$, and show that they are equal.  To
define $\tind$, let $M\hraw\R^k$ be a smooth embedding, defining a
normal bundle $\nu(M)\raw M$ and associated pushforwards $T(M)\hraw
T(\R^k)$ and $T(\nu(M))\raw T(M)$.  Since the latter bundle has a
complex structure (or, more generally, is even-dimensional and
K-oriented), one has the K-theory Thom isomorphism $\ta:K^0(T(M))\raw
K^0(T(\nu(M)))$.  Identifying $T(\nu(M))$ with a tubular neighbourhood
of $T(M)$ in $T(\R^k)$, one has $T(\nu(M))\hraw T(\R^k)$ as an open
set, so that one has a natural extension map $\ps:K^0(T(\nu(M)))\raw
K^0(T(\R^k))$.  Finally, for $T(\R^k)=\R^{2k}$ one has the Bott
isomorphism $\bt_k:K^0(\R^{2k})\raw\Z$. Identifying
$T(M)$ with $T^*(M)$ through some metric, $\tind$ is the composition
\begin{equation}
\tind=\bt_k\circ\ps\circ\ta: K^0(T^*(M))\raw\mathbb{Z}. \label{deftopi}
\end{equation}
Using some algebraic topology, it is easy to show that
\begin{equation}
\tind(x)=(-1)^{\dim(M)}\int_{T^*(M)}\mathrm{ch}(x)\wedge\pi^*\mathrm{td}(T^*(M)\ot\C),
\label{algtop}
\end{equation}
where $\mathrm{ch}:K^0(T^*(M))\raw H_c^*(T^*(M))$ is the Chern
character, $\pi:T^*(M)\raw M$ is the canonical projection, and
$\mathrm{td}(E)\in H^*(M)$ is the Todd genus of a complex vector
bundle $E\raw M$. 

The analytic index $\aind:K^0(T^*(M))\raw\mathbb{Z}$ is defined by
\begin{equation}
\aind(\sg_P)=\mathrm{index}(P). \label{defaind}
\end{equation}
Here $P:\cin(E)\raw\cin(F)$ is an elliptic
pseudodifferential operator between complex vector bundles  $E$ and $F$ 
over  $M$, with principal symbol $\sg_P\in  K^0(T^*(M))$, and
\begin{equation}
\mathrm{index}(P)=\dim\, \ker (P)-\dim\, \mathrm{coker}(P).
\end{equation}
 Atiyah and Singer \cite{AS1} formulate two axioms which $\tind$ 
is trivially shown to satisfy, and which uniquely characterize $\tind$ 
as a map from $K^0(T^*(M))$ to $\mathbb{Z}$. 
The burden of their proof of the index theorem in K-theory
\begin{equation}
\tind=\aind \label{Kind}
\end{equation}
is to show that $\aind$ satisfies these axioms as well. Combining 
(\ref{algtop}), (\ref{defaind}), and (\ref{Kind}), one then obtains the usual cohomological
form of the index theorem \cite{AS3}, viz.\ 
\begin{equation}
\mathrm{index}(P)=(-1)^{\dim(M)}\int_{T^*(M)}\mathrm{ch}(\sg_P)
\wedge\pi^*\mathrm{td}(T^*(M)\ot\C).\label{indextheorem}
\end{equation}

This proof has a number of drawbacks. It is not easy to show that (\ref{defaind})
is well defined; one must establish that $\mathrm{index}(P)$ only depends on the
symbol  class $\sg_P$, and that $K^0(T^*(M))$ is exhausted by elements of that form.
Furthermore, the definition of $\tind$ looks artificial. All in all, it would seem preferable
to have natural map
\begin{equation}
\qind: K^0(T^*(M))\raw\mathbb{Z}
\end{equation}
 to begin with, and to show
that this a priori defined map satisfies both
\begin{equation}
\qind(x)=(-1)^{\dim(M)}\int_{T^*(M)}\mathrm{ch}(x)\wedge\pi^*\mathrm{td}(T^*(M)\ot\C)
\label{qtind}
\end{equation}
and
\begin{equation}
\qind(\sg_P)=\mathrm{index}(P). \label{qaind}
\end{equation}
This would immediately imply (\ref{indextheorem}).

This program may  indeed be realized \cite{Con,ENN2,Hig,Trout}. We start from the continuous field
(\ref{CF1}), defining the map (\ref{Kmap}). Composing this map with
 the trace $\mathrm{tr}: K_0(B_0(L^2(M)))\stackrel{\raw}{\cong}\mathbb{Z}$), one may put
\begin{equation}
\qind=\mathrm{tr}\circ\CQ_0.  \label{inda}
\end{equation}
Connes (cf.\ Lemma II.5.6 in \cite{Con}) claims that this map
coincides with $\aind$, which is true, but this equality actually
comprises half of the proof of the index theorem!  The computations
establishing (\ref{qaind}) may be  found in \cite{ENN2,Hig,LMN,Trout}.

One way to prove (\ref{qtind}) is to
note that the continuous field (\ref{CF1}) extends to a continuous
field defined by $A_X=A\ot C(X)$, where $X$ is any compact Hausdorff
space (cf.\ Thm.\ 2.4 in \cite{ENN1}).  Using Cor.\ 3.2 in
\cite{ENN1}, the associated maps (\ref{Kmap}) $\CQ_*^X: K^*(T^*(M)\x
X)\raw K^*(X)$ are easily seen to be natural in $X$, and to be 
homomorphisms of $K(X)$ modules. Furthermore, a lengthy calculation
given in \cite{Hig} shows that $\CQ_0^M(\lm_M)=1$, where $\lm_M\in
K^0(M\x T^*(M))$ is a generalized Bott element defined in \cite{Hig}.
As shown in \cite{Hig}, by a straightforward topological argument
these three properties imply (\ref{qtind}). Also see \cite{Trout} for a
different proof. 

Another approach to proving (\ref{qtind}) is due to Connes; see \S
II.5 of \cite{Con}. First, extend the embedding $M\hraw\R^k$ to
$j:M\hraw\R^{2k}$ by mapping $x\in\R^k$ to $(x,0)\in\R^{2k}$.  Recall
that $G_M$ is the tangent groupoid of $M$, with base $G_M^{(0)}=I\x
M$. Now
\begin{equation}
 P=G_M^{(0)}\x\R^{2k} \label{defEGM}
\end{equation} 
is a right $G_M$ space through the obvious map $P\stackrel{\al}{\raw} G_M^{(0)}$,
i.e., $\al(u,X)=u$, and the action is
given by
\begin{equation} (r(\gm),X)\gm=(s(\gm), X+ h(\gm)). \label{Xh}
\end{equation}
Here $h(\hbar=0,\xi_x)=j_*(\xi_x)$ and $h(\hbar,x,y)=(j(x)-j(y))/\hbar$.
 This action defines
the semidirect product groupoid $P\rtimes G_M$.
From (\ref{sr})  and (\ref{Xh}) one reads off the source and range 
projections $s,r:P\rtimes G_M\raw P$ as 
\begin{eqnarray}
s(r_{G_M}(\gm),X,\gm) & = & (s_{G_M}(\gm), X+h(\gm)); \nn \\
r(r_{G_M}(\gm),X,\gm) & = & (r_{G_M}(\gm), X).
\end{eqnarray}

 Connes's first
observation (Prop.\ 7 on p.\ 104 of \cite{Con}) is that
\begin{equation}
K^*(G_M)\cong K^*(P\rtimes G_M). \label{Con1}
\end{equation}
This follows, because $C^*(P\rtimes G_M)\cong C^*(G_M)\rtimes \R^{2k}$
with respect to a suitable action of $\R^{2k}$ on $C^*(G_M)$, This isomorphism is easily established by a Fourier transformation on the
$\R^{2k}$ variable, and implies (\ref{Con1}) by Connes's Thom
isomorphism \cite{Bla,Con,ENN1}. It is interesting to regard (\ref{Con1})
as a proof of the Baum--Connes conjecture for $G_M$. Indeed, 
one may take 
\begin{equation}
\underline{E}G_M=G_M^{(0)}\x\R^{2k};
\end{equation}
in particular, the $G_M$ action on $P$ is free and proper. As a groupoid and as a $G_M$ space, $T_{G_M}(P)$ is just $P\x \R^{2k}$ over $P\x \{0\}$,
 i.e., the  direct product of $P$ as a space with the given $G_M$ action and
$\R^{2k}$ as an abelian group with the trivial $G_M$ action. Therefore,
\begin{equation}
C^*(T_{G_M}(P)\rtimes G_M)\cong C^*(P\rtimes G_M)\ot C_0(\R^{2k}),
\end{equation}
 so that,  using (\ref{Ktop}) and Bott periodicity, one has
\begin{equation}
K^*_{\mathrm{top}}(G_M) \cong K^*(P\rtimes G_M). \label{Ktop1}
\end{equation}
The analytic assembly map $\mu:K^*_{\mathrm{top}}(G_M)\raw K^*_r(G_M)$
is precisely the map occurring in Connes's Thom
isomorphism. Note that $K^*_r(G_M)=K^*(G_M)$, both being isomorphic
to $K^*(T^*(M))$. 

The second main observation \cite{Con} is that
the Lie groupoid $P\rtimes G_M$ is Morita equivalent to the space
\begin{equation} 
\underline{B}G_M=\{0\}\x T(\nu(M)) ) \cup ((0,1]\x\R^{2k}.
\end{equation}
Here $T(\nu(M))=\nu(M)\x \R^k$
is actually the normal bundle of $M\stackrel{j}{\raw}\R^{2k}$, so 
this is a special case of (\ref{GMN}). 

Looking separately at the cases $\hbar=0$ and $\hbar>0$, 
it is easily seen that $\underline{B}G_M$ is diffeomorphic to the orbit space 
$P/G_M$ of the $G_M$ action on $P$ (which also explains the notation, as $P=\underline{E}G_M$). 
This coincides with the orbit space  $P/(P\rtimes G_M)$ of the
$P\rtimes G_M$ action on its own base space $P$, which is free and 
proper. The orbit space $\underline{B}G_M$ acts trivially on $P$, and it follows that $P$ is a $(\underline{B}G_M,P\rtimes G_M)$ equivalence
\cite{MRW}. Hence $\underline{B}G_M$ and $P\rtimes G_M$ 
are Morita equivalent.

It follows that
\begin{equation}
K^*(P\rtimes G_M)\cong K^*(\underline{B}G_M), \label{Con2}
\end{equation}
and hence, by (\ref{Con1}), 
\begin{equation}
K^*(G_M)\cong K^*(\underline{B}G_M). \label{Con3}
\end{equation}

Now both $C^*(G_M)$ and $C^*(\underline{B}G_M)=C_0(\underline{B}G_M)$
are $C(I)$ \ca s,  defining continuous fields by
Lemma \ref{Ramlem}. We decorate maps associated to the second field
with a hat. For example, the associated 
 maps (\ref{Kmap}) are $\CQ_*: K^*(T^*(M))\raw K_*(B_0(L^2(M)))$ and
$\hat{\CQ}_*: K^*(T(\nu(M)))\raw K^*(\R^{2k})$, respectively. 
We have already dealt with $\CQ_0$; 
it is easily seen that $\hat{\CQ}_0$ is the extension map $\ps$.
The isomorphism (\ref{Con3}), which we call $\al^*$, induces
isomorphisms $\al^*_{\hbar}: K_*(C^*(G_M)_{\hbar})\raw 
K^*(C_0(\underline{B}G_M)_{\hbar})$ such that
$\al^*_{\hbar}\circ \pi_{\hbar}=\hat{\pi}_{\hbar}\circ\al^*$,
 for any $\hbar\in I$. It can be checked that $\al^0_0:
K^0(T^*(M))\raw K^0(T(\nu(M)))$ is the Thom isomorphism $\ta$, and
that $\al^0_1:K_0(B_0(L^2(M)))\raw K^0(\R^{2k})$ is $\bt_k\inv\circ\mathrm{tr}$. It follows from the definition of 
$\CQ_*$ and $\al_{\hbar}$ that one has $\al_1^*\circ \CQ_*=
\hat{\CQ}_*\circ\al_0^*$. Using (\ref{deftopi}) and (\ref{inda}),
the last equality with $*=0$ immediately implies 
$\aind=\tind$, and hence (\ref{qtind}) from (\ref{algtop}). 
This proof of the index theorem has great conceptual beauty.

We close with some comments on the Baum--Connes conjecture in E-theory
in the light of the above considerations.
For $M=\mathbb{R}^k$, the map
$\qind:K^0(\mathbb{R}^{2k})\raw\mathbb{Z}$ is the inverse of the Bott
map, so that Atiyah's index theory proof of the Bott periodicity
theorem \cite{At1} may actually be rewritten in terms of deformation
quantization \cite{ENN1,Trout,GBV}.\footnote{Note that the localization to $[0,1]$ of the continuous field associated to the Heisenberg group used in
\cite{ENN1} is the same as the field defined by $C^*(G_{\R^k})$, cf.\
\S II.2.6 of \cite{Lan}, so that the approach in
\cite{ENN1} is really based on deformation quantization.}
The fact that the ``classical algebra'' $C_0(\mathbb{R}^{2k})$ and
the ``quantum algebra'' $B_0(L^2(\R^k))$ have the same
K-theory is peculiar to this special case; for general $M$
this will, of course, fail.  Indeed, the Baum--Connes conjecture
mau be seen as a test of the rigidity of K-theory under 
deformation quantization.
Connes's
interpretation of the Baum--Connes conjecture as a $G$ equivariant
version of Bott periodicity \cite[\S II.10.$\ep$]{Con} is consistent
with this picture, since the field
(\ref{CF3}) underlying the Baum--Connes conjecture is just a $G$
equivariant version of the field (\ref{CF1}).  
\begin{footnotesize}

\end{footnotesize}
\end{document}